TEL AVIV UNIVERSITY

# Frame-based codes for partially active NOMA

By

Maya Slamovich

Supervisor:

Prof. Ram Zamir

# Abstract

Non-orthogonal multiple-access (NOMA) is a leading technology which gain a lot of interest this past several years. It enables larger user density and therefore is suited for modern systems such as 5G and IoT.

In this paper we examined different frame-based codes for a partially active NOMA system. It is a more realistic setting where only part of the users, in an overly populated system, are active simultaneously. We introduce a new analysis approach were the active user ratio, a system's feature, is kept constant and different sized frames are employed. The frame types were partially derived from previous papers on the subject [1] [2] and partially novel such as the LPF and the Steiner ETF. We learned the best capacity achieving frame depends on the active user ratio and three distinct ranges where defined.

In addition, we introduced a measure called practical capacity which is the maximal rate achieved by simple coding scheme. ETF always achieves the best practical capacity while LPF and sparse frame are worse than a random one.

## Contents

# 1. Introduction

Up until recent years Orthogonal multiple-access (OMA) was the optimal transmission design for multiple-access communications. It achieves the ultimate total throughput in a fully loaded Gaussian channel, where the number of users equals the total number of available orthogonal resources. These resources can be time slots in a TDMA system, frequency bands in an OFDMA system, code sequence in a CDMA system and more.

With the rising demand for larger user density in current technologies such as IoT and 5G, the non-orthogonal multiple access (NOMA) scheme is gaining popularity. The NOMA scheme targets overloaded regimes, where the number of potential users exceeds the number of available resources. NOMA can be roughly split into two domains – power and code. This paper will examine the latter.

In our paper we consider a partially active NOMA system for which different transmission schemes will be evaluated. We start with a fully active NOMA system where the signals of N users are multiplexed over M shared orthogonal resources:

$$\underline{y} = c \cdot \underline{\underline{F}}\,\underline{x} + \underline{n}$$

X is a N-dimensional complex vector comprising the coded symbols of the users. Assuming no cooperation between encoders and gaussian signaling, the input vector x in distributed as $\underline{x} \sim CN(0, I_N)$. $\underline{n}$ denotes the n-dimensional complex additive white gaussian noise vector $\underline{n} \sim CN(0, I_M)$. F denotes the MxN user-resource mapping (user n occupies resource m if $F_{m,\text{n}} \neq 0$. c is a normalization factor which is set so that:

$$F = \begin{pmatrix} f_{M\times 1}^{1} & f_{M\times 1}^{2} & \cdots & f_{M\times 1}^{N} \end{pmatrix}$$
$$\mathrm{E}\left[ \left\| c \cdot f_{M\times 1}^{i} \cdot x_i \right\|^2 \right] = SNR, \quad \forall i = 1 \ldots N$$

The parameter SNR thus designates the received signal to noise ratio of each user. In addition, we define $\gamma^{-1} \triangleq \dfrac{N}{M} \geq 1$ as the system load (resources per user).

Now we can generalize the model to fit a partially active NOMA system which is a more realistic setting where not all users are active simultaneously. The active users are randomly selected according to Bernoulli(p) distribution. This means we are only interested in a random subset of the frame vectors, or columns of F. In other words, each of the vectors $f_{M\times 1}^{1} \quad f_{M\times 1}^{2} \quad \cdots \quad f_{M\times 1}^{N}$ is erased from the frame with probability $1-p$.
We define the subframe as:

$$\tilde{\underline{\underline{F}}}_{M\times|\kappa|} = \left\{ f_{M\times 1}^{i} \right\}_{i\in\kappa}$$
$$\mathrm{E}\left[ |\kappa| \right] = pN \in \mathbb{N} \leq N$$

However, for model simplicity, we chose a constant number of active users $K = p \cdot N \in \mathbb{N}$.
The active users are chosen uniformly at random from $\{1, \ldots, N\}$.
So in total the random subframe is defined as:

$$\tilde{\underline{\underline{F}}}_{M\times K} = \left\{ f^{i}_{M\times 1} \right\}_{i\in\kappa}$$
$$\left|\kappa\right| = K = pN \in \mathbb{N} \le N$$

Notice that the fully active NOMA system is equivalent to a partially active one with p=1.

we define $\beta \triangleq \frac{K}{M} = \frac{pN}{M} \le 1$ as the system's active load (number of active users per resource).

Throughout this paper we evaluate and compare four different mapping approaches ( $\underline{\underline{F}}$ frame) between users and resources:

1. Regular sparse mapping [2] – sparse spreading codes comprising only a small number of non-zero elements. Its main advantage is its inherent receiver complexity reduction. The mapping is dubbed regular when each user occupies a fixed number of resources and each resource is used by a fixed number of users, i.e. each column of $F$ contains a constant $d \in \mathbb{N}^{+}$ number of nonzero values and each row $\gamma^{-1}d \in \mathbb{N}^{+}$ nonzero values. Irregular is when the respective numbers are random and only fixed on average. Regular mapping is superior to irregular mapping so we will focus on the former.
2. Low pass frame (LPF) – A type of unit-norm tight frame (UNTF) which is constructed by taking the upper $M\times N$ section of a $N\times N$ DFT matrix. Mapping using a UNTF was proven analytically to maximize the channel capacity for the fully active (p=1) case [3]. More on the frames structure in the next chapter.
3. Equiangular tight frame (ETF) – ETFs are the closest equivalent of orthonormal bases in terms of incoherence. All distinct pairs of column vectors of $F$ have the same cross correlation $\left|\left\langle f^{i}, f^{j}\right\rangle\right| = const$ . ETFs where shown to minimize the d-th order moments of a partially active setting [4]. More on the frames structure in the next chapter.
4. Gaussian i.i.d. Frame – a frame which has i.i.d (independent and identically distributed) normal entries with zero mean and $1/M$ variance.

We will start with a review on frame theory, different characteristics and importance. Then we will present our frame generation algorithms and elaborate on the system's analysis and results. We will conclude with final discussions and future studies to be performed. All codes are attached as appendixes at the end of this document.

## 2. Frame Theory

The general focus of Frame Theory is the study of overcomplete representations of orthonormal bases. Its theorems and conclusions can be employed to better understanding of our mapping problem.

A matrix $F \in \mathbb{C}^{M \times N}$ with $M \le N$ is a frame over the finite Hilbert space $\mathbb{C}^M$ if its linearly dependent column vectors $\{f_n\}_{n=1}^{N} \in \mathbb{C}^M$ satisfy:

$$a\|y\|^2 \le \sum_{n=1}^{N} |\langle f_n, y\rangle|^2 \le b\|y\|^2, \quad \forall y \in \mathbb{C}^M$$

Where the finite coefficients $0 < a < b$ are referred to as the frame bounds.

A frame is said to be *tight* if $a = b$ leaving us with:

$$\sum_{n=1}^{N} |\langle f_n, y\rangle|^2 = a\|y\|^2, \quad \forall y \in \mathbb{C}^M$$

Which is equivalent to a frame satisfying:

$$FF^H = aI_M$$

A tight frame is called a UNTF when all vectors $\{f_n\}_{n=1}^{N} \in \mathbb{C}^M$ are unit-norm, which further implicate that $a = b = \frac{N}{M} = \gamma^{-1}$ meaning:

*Equation 1*

$$FF^H \overset{UNTF}{=} \frac{N}{M} I_M$$

Now we define the vectors' cross correlation:

$$c_{n,k} \triangleq \langle f_n, f_k\rangle = f_n' f_k = \sum_{i=1}^{M} F_{i,n}^* F_{i,k}$$

Where for UNTF:

$$c_{i,i} = \|f_i\|^2 = 1$$

The Welch Bound lower bounds the mean-square (ms) cross correlation [5]:

$$I_{ms}(F) \triangleq \frac{1}{(N-1)N} \sum_{n \ne k}^{N} |c_{n,k}|^2 \ge \frac{N-M}{(N-1)M}$$

And is achieved with equality iff F is a UNTF.

The Welch Bound implies a bound on the maximum-square cross correlation [5]:

*2 Equation*

$$I_{\max}(F) \triangleq \max_{1 \le n \le k \le N} \left| c_{n,k} \right|^2 \ge \frac{N-M}{(N-1)M}$$

We define an Equiangular Tight Frame (ETF) as a UNTF which satisfies:

$$\left| c_{n,k} \right|^2 = const = \frac{N-M}{(N-1)M} \qquad \forall n \ne k$$

An ETF, as explained before, is the closest equivalent of orthonormal bases in terms of incoherence. Meaning, it's a UNTF with the maximal spatial angular displacement of the frame vectors pairs $(f_n, f_k), \quad n \ne k$. Which means ETF achieves the max-WB with equality.

The existence of ETF is limited to particular dimensions pairs $(M, N)$ and is not easily generated.

# 3. Frame Generation

## 3.1. Sparse Frame

Sparse frames were constructed according to the description in [2]. The frame is of size MxN with exactly $2 \le d \in \mathbb{N}^+ < \infty$ non-zero entries in each column, and $2 \le \gamma^{-1} d \in \mathbb{N}^+ < \infty$ non-zero entries in each row . Each non-zero element is uniformly distributed over the unit circle in $\mathbb{C}$. Meaning $\gamma^{-1} d$ is limited to normal numbers only. Code can be found on appendix 8.1.

In order to verify our implementation, we compared the eigenvalues' pdf of $\frac{1}{d} F F^H$, calculated using our implementation, to the theoretical one (given in [2]). Looking at Figure 1 you can see we got a good match.

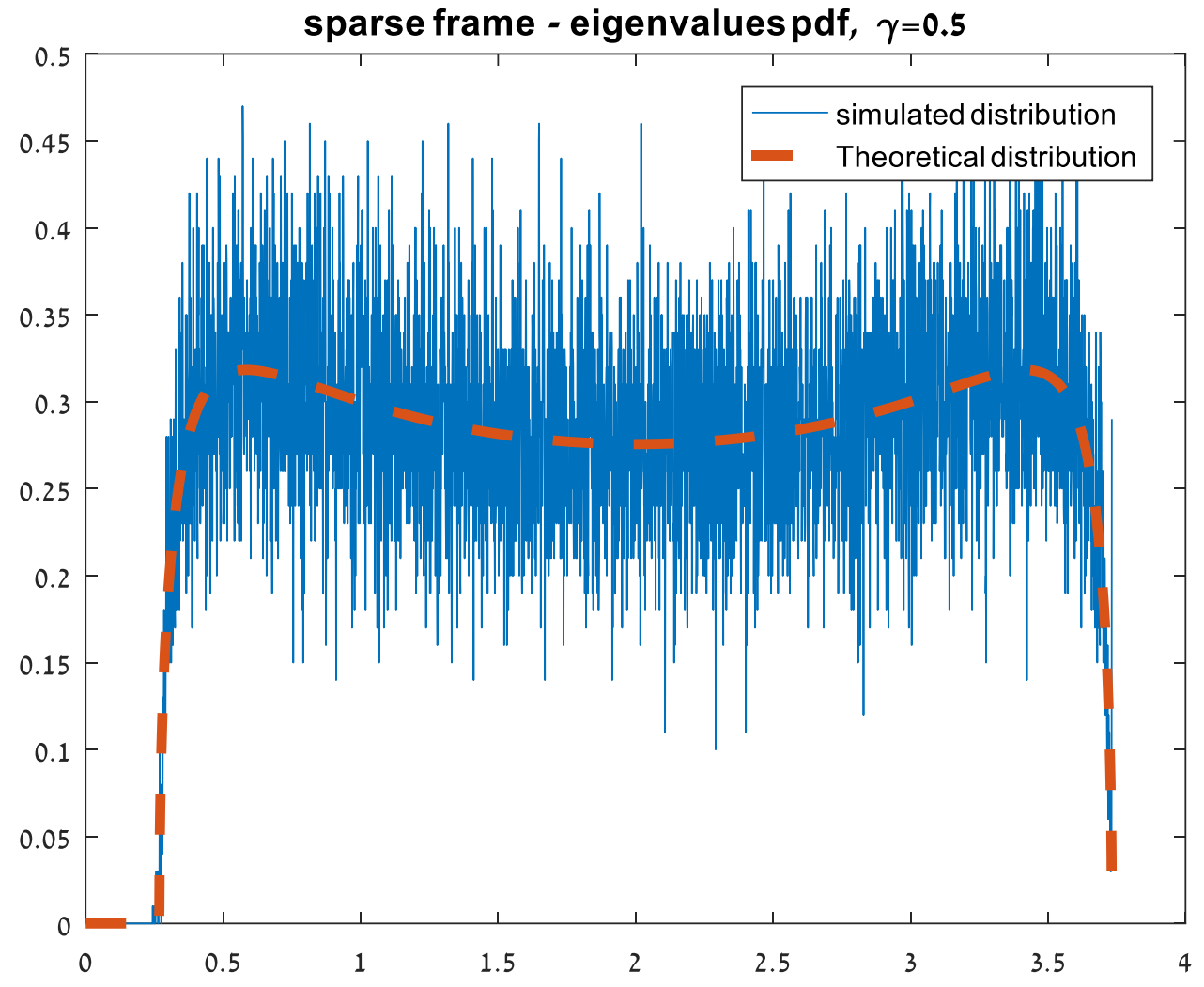

*Figure 1- comparison between thoretical sparse matrix eigenvalues' pdf to the simulated one*

## 3.2. LPF – Low Pass Frame

A LPF construction is relatively easy. You simply create a DFT matrix of size N and erase the last N-M rows. Afterwards you normalize each column to unit norm. In addition, we verified the frame satisfies Equation 1. Code can be found on appendix 8.2.

## 3.3. ETF – Equiangular Tight Frame

As explained before, an ETF is a frame with normalized vectors that achieves the Max Welch bound. Meaning that the inner product of its vectors equals $\sqrt{\frac{N-M}{M(N-1)}}$.

An ETF does not exist for every (M,N) combination but rather only for specific ones. Over the years several methods of construction where developed. In this paper we used two of them – the difference sets method and Steiner method.

### 3.3.1. Difference sets

A difference set $D$ is a collection of $k$ distinct residues $d_1, \ldots, d_k \bmod v$, for which the congruence $d_i - d_j = b (\bmod v)$ has exactly $\lambda$ distinct solution pairs $d_i, d_j$ in $D$ for every $b \neq 0(\mathrm{mod}\, v)$. Each difference set is defined by the values $(v,k,\lambda)$. For example a difference set $(7,3,1)$ can be the group $\{0,1,3\}$ because we can get any number in $(0,1,2,3,4,5,6) \mathrm{mod}\, 7$ by substituting one pair in the set.

It was shown in [6] that by collecting rows from a $N \times N$ DFT matrix which are indexed by $D$ you can construct an $M \times N$ ETF if and only if $D$ is a difference set with $v = N$ and $k = M$. This means we can construct ETFs rather easily by using difference sets.

There are several known families of difference sets. In our paper we used the Paley family which is $\left(p^i+1, \frac{p^i+1}{2}, \lambda\right)$ p^i provided that $p$ is a prime number and $i \geq 1$. Using this family, we constructed ETFs with $\gamma = \frac{M}{N} = \frac{1}{2}$.

The construction of ETF using the Paley difference set method has low complexity and can ber computed using Matlab.

### 3.3.2. Steiner equiangular tight frames

In [7] a new method for constructing ETFs was presented. The steps are as follows:

1. It start with the transposed incidence matrix of a $(2,k,v)$- Steiner system which has the following characteristics:
   a. $v$ columns
   b. $b = \frac{v(v-1)}{k(k-1)}$ rows
   c. every row contains $k$ elements
   d. every column contains $r = \frac{v-1}{k-1}$ elements
   e. any pair of elements is contained in exactly one row.

   For example – a $(2,2,4)$ - Steiner system whose transposed incidence matrix is:

$$A^T = \begin{bmatrix} + & + & & \\ + & & + & \\ + & & & + \\ & + & + & \\ & + & & + \\ & & + & + \end{bmatrix}$$

   This is a matrix with ones and zeros (+ implies one).
2. Next, construct a matrix with unimodular entries (absolute value equals one) and orthogonal rows of size $\frac{N}{v} \times \frac{N}{v}$ such as a Hadamard matrix or a DFT matrix.
   For example – for $N = 16$ we can choose a Hadamard matrix:

$$H = \begin{bmatrix} + & + & + & + \\ + & - & + & - \\ + & + & - & - \\ + & - & - & + \end{bmatrix}$$

3. To form $F$ , in each column of $A^T$ we replace each 1-valued entry with a distinct row of H and normalize the columns. One may choose a different sequence of rows of $H$ for each column.
   For example – always choosing the second, third and fourth rows yields an $6 \times 16$ ETF:

$$F = \frac{1}{\sqrt{3}} \begin{bmatrix} + & - & + & - & + & - & + & - & & & & & & & & \\ + & + & - & - & & & & & + & - & + & - & & & & \\ + & - & - & + & & & & & & & & & + & - & + & - \\ & & & & + & + & - & - & + & + & - & - & & & & \\ & & & & + & - & - & + & & & & & + & + & - & - \\ & & & & & & & & + & - & - & + & + & - & - & + \end{bmatrix}$$

   You can verify the rows of $F$ are orthogonal and have constant norm which equals to $\frac{N}{M}$ , implying $F$ is a UNTF. One can also easily see that the inner products of two columns from the same block are $-\frac{1}{3}$, while the inner products of columns from distinct blocks are $\pm\frac{1}{3}$, giving us in total an ETF frame.

A recursive code was written in order to construct the $A^T$ matrixes used in the Steiner method. On each function call a row of $A^T$ was populated by one of the combinations available (a combination that doesn't violate any of the matrix's requirements). If a row can't be populated without violating the requirements a fail message is returned so the previous call will choose a different population option from the ones possible. Code can be found on appendix 8.3.

Unfortunately, the computing complexity of the code was quite high and would have taken several days, if not weeks, to finish, especially for the larger matrixes. As a result, I decided to apply parallel computing and use Google Cloud's multi-core machines. This version of the code can be found on appendix 8.4 . A log of the machine's CPU usage is given in Figure 2 so you could see that for really large matrixes the code run for a whole two weeks even on the cloud.

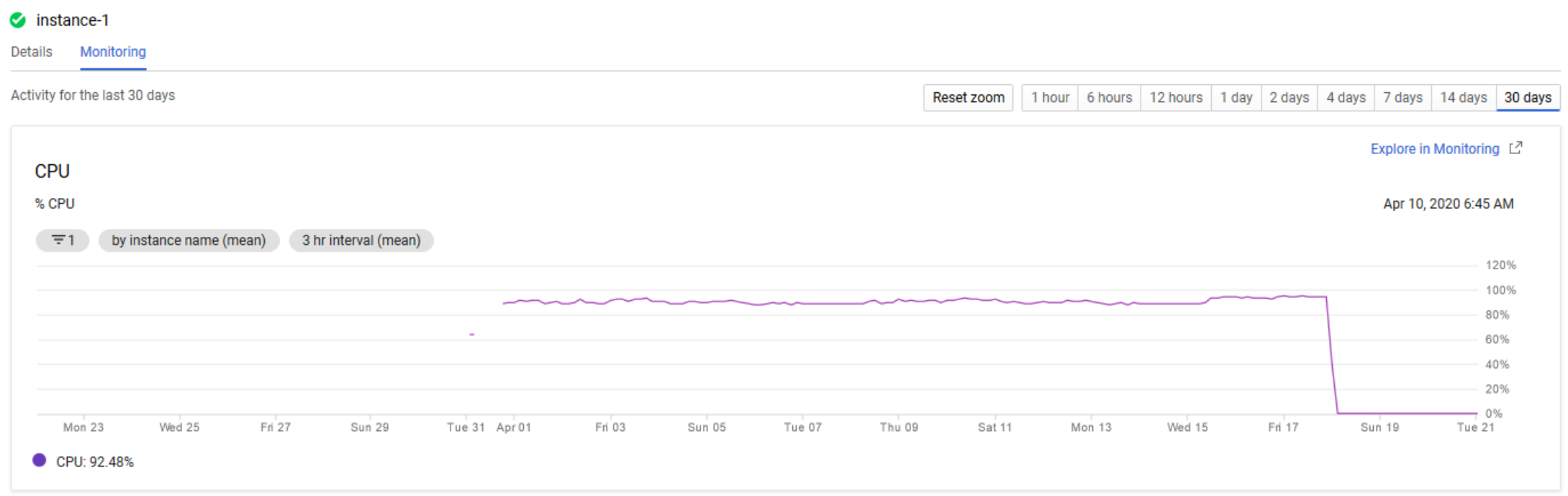

*Figure 2 - cloud machine's CPU log*

### 3.3.3. Verification

It was proven that the pdf of the eigenvalues of a random subset of an ETF is equal to the Manova distribution [1]. We will verify this fact in order to check our ETF's construction process.

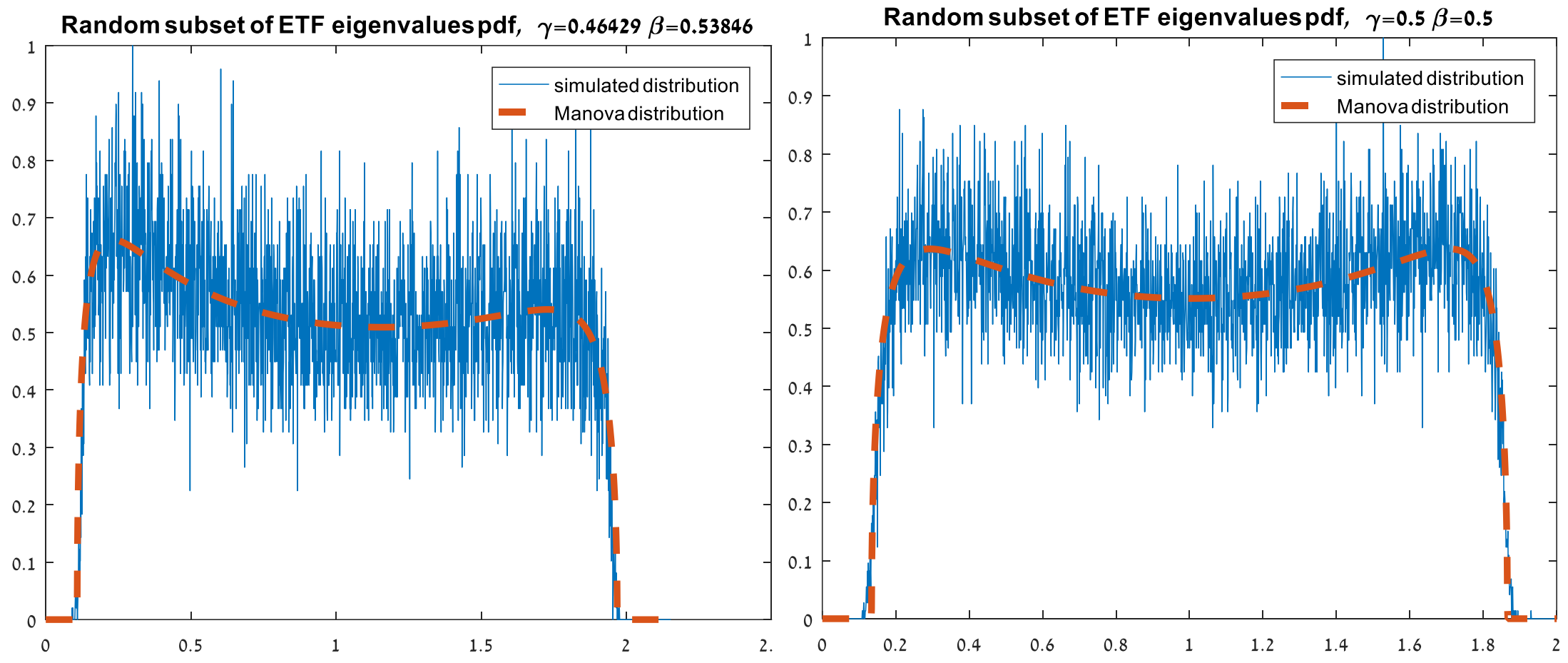

*Figure 3 - comparison between random subset of ETF's eigenvalues' pdf to the Manova distribution*

You can see from Figure 3 we got a good match between the distributions so we can feel confident about our ETF implementation.

## 3.4. Gaussian i.i.d. Frame

As explained before, this is a frame which has i.i.d (independent and identically distributed) normal entries with mean zero and variance $1/M$ . The chosen variance guarantees unit normalized vectors, meaning:

$$\mathrm{E}\left[\|f_i\|^2\right]=1$$

$$\|f_i\|^2 \xrightarrow[M,N\to\infty]{} 1$$

The construction is straight forward and is given in 8.5.

In order to verify our implementation, we compared the eigenvalues' pdf of a random subset of the frame to the Marcenko-Pastur density [1]. Looking at Figure 4 you can see we got a good match.

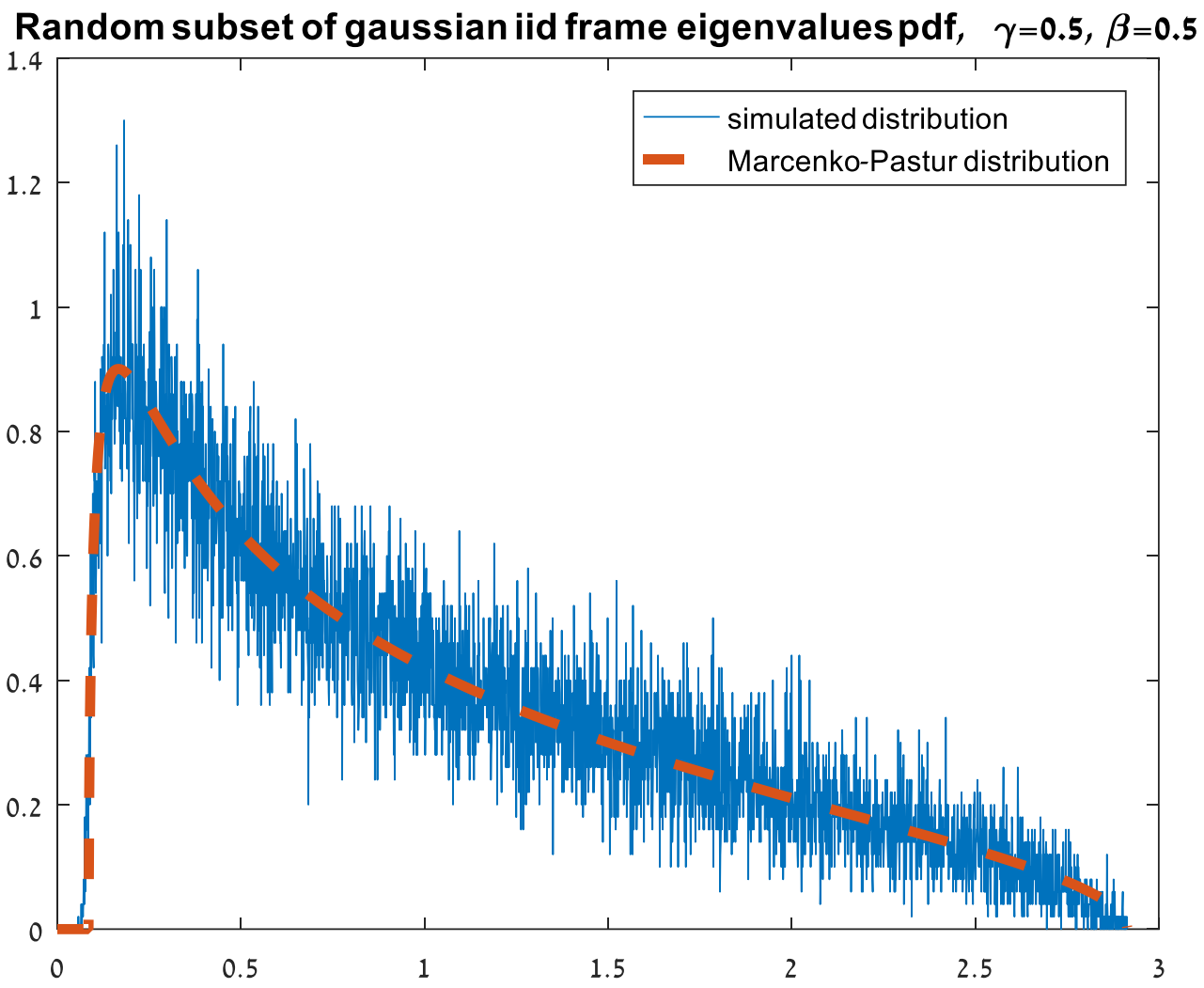

*Figure 4 - comparison between random subset of gaussian iid frame's eigenvalues' pdf to the Marcenko-Pastur distribution*

# 4. Performance measures

In order to evaluate and compare the different frames' performance we used four different measures which can be divided into two main types:

1. Capacity – the channel capacity of a linear Gaussian channel was proven to be [8]:

$$C[bits/s] = \mathrm{E}\left\{\log_2\left(\det\left(I + SNR \cdot \tilde{F}^H \tilde{F}\right)\right)\right\}$$

   The capacity can be re-written as a function of the eigenvalues:

   *Equation 3*

$$C[bits/s] = \mathrm{E}\left\{\log_2\left(\det\left(I + SNR \cdot \tilde{F}^H \tilde{F}\right)\right)\right\} =$$
$$= \mathrm{E}\left\{\log_2\left(\prod_{i=1}^{K}(1 + SNR \cdot \lambda_i)\right)\right\} = \mathrm{E}\left\{\sum_{i=1}^{K}\log_2(1 + SNR \cdot \lambda_i)\right\}$$

   Where $\lambda_i$ are the eigenvalues of $\tilde{F}\tilde{F}^H$.

2. Practical capacity – It's well known that the system capacity is the higher bound on the achievable rate of a system. Achieving it will most likely require extremely sophisticated and complex coding schemes. For simple systems, we introduce a more practical bound, achievable using simple schemes, which we call the practical capacity. It will also help us visualize and evaluate the effect of eigenvalues and SNR on the system. Practical capacity is defined as:

   *Equation 4*

$$C_p[bits/s] = \mathrm{E}\left\{\log_2\left(\det\left(SNR \cdot \tilde{F}^H \tilde{F}\right)\right)\right\}$$

   And as a function of the eigenvalues:

$$C_p[bits/s] = \mathrm{E}\left\{\sum_{i=1}^{K}\log_2(SNR \cdot \lambda_i)\right\} = K\log_2(SNR) + \mathrm{E}\left\{\sum_{i=1}^{K}\log_2(\lambda_i)\right\} =$$
$$= K\log_2(SNR) + \mathrm{E}\left\{\log_2\left(\prod_{i=1}^{K}\lambda_i\right)\right\}$$

Each of the measure types above can be normalized per user or per resource giving us a total of four measures:

1. Capacity per user:

$$C[bits/s/\mathrm{user}] = \frac{1}{K}\log_2\left(\det\left(I + SNR \cdot \tilde{F}^H \tilde{F}\right)\right)$$

2. Capacity per resource – also known as spectral efficiency when the resources are frequency bands:

   *Equation 5*

$$C[bits/s/\mathrm{resource}] = \frac{1}{M}\log_2\left(\det\left(I + SNR \cdot \tilde{F}^H \tilde{F}\right)\right) =$$
$$= \frac{1}{M}\frac{K}{K}\log_2\left(\det\left(I + SNR \cdot \tilde{F}^H \tilde{F}\right)\right) = \beta \cdot C[bits/s/user]$$

3. Practical capacity per user:

$$C_p\left[bits/s/\mathrm{user}\right]=\frac{1}{K}\log_2\left(\det\left(SNR\cdot\tilde{F}^H\tilde{F}\right)\right)$$

4. Practical capacity per resource:

$$C_p\left[bits/s/\mathrm{resource}\right]=\beta\cdot C_p\left[bits/s/user\right]$$

# 5. Analysis

### 5.1.1. Fully active NOMA system

On [2] they showed the superiority of sparsed frames over different allocation techniques regarding capacity per resource for classic NOMA systems. In our paper we first concentrated on reproducing those results and comparing them to additional frame structures.

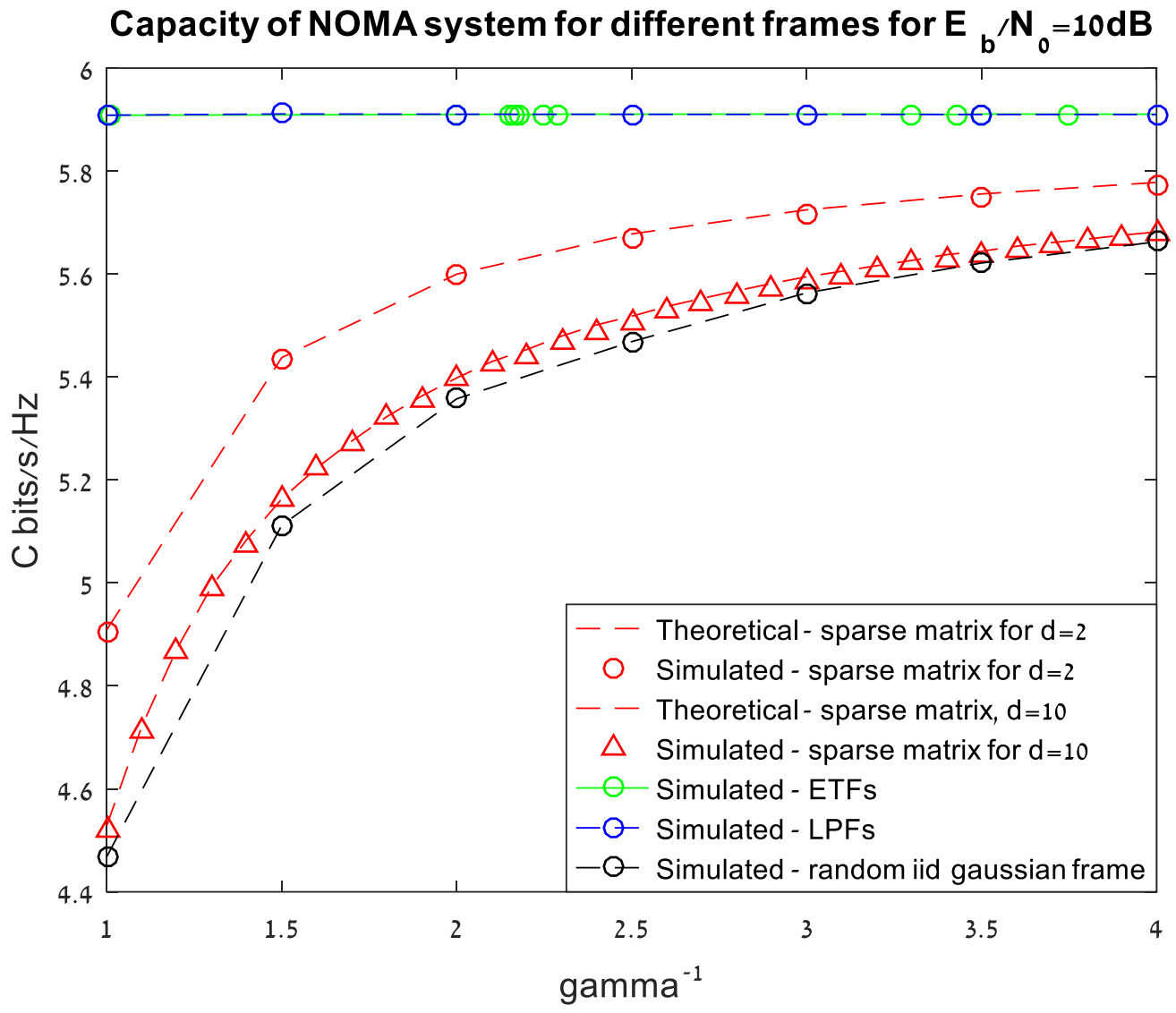

*Figure 5 - [1] results reprodution for fully active NOMA system with additional fames*

The reconstruction was successful, as can be seen in Figure 5 . The sparser the frame, the higher the capacity per resource and its obviously superior to the random one. On the other hand, the sparser the frame the less $\gamma$ values we can choose as $d\gamma^{-1}$ must be a natural number. For example if $d = 10$ $\gamma^{-1}$ can take any of the following values $\gamma^{-1} \in \{1, 1.1, 1.2, ...\}$ where as if $d = 2$ we are left with fewer options $\gamma^{-1} \in \{1, 1.5, 2 ...\}$.

It can be seen that a UNTF (uniform tight frame) of any kind, achieves higher capacity and is the best mapping approach from the ones examined. Choosing an ETF isn't worthwhile since it shows no improvement over a simple UNTF.

It's interesting that the capacity when using an UNTF is not affected by $\gamma$. The reason for that is that the simulation was made for constant $\frac{E_b}{N_0}$ which cancels out the $\gamma$ effect as can be seen here:

$$C = \frac{1}{M}\log_2\left(\det\left(I + SNR \cdot FF^H\right)\right) \underset{F \text{ is } UTF}{=} \frac{1}{M}\log_2\left(\det\left(I + \frac{SNR}{\gamma}\cdot I\right)\right) =$$

$$= \frac{1}{M}\log_2\left(\left(1+\frac{SNR}{\gamma}\right)^M\right) = \log_2\left(1+\frac{SNR}{\gamma}\right) \underset{\frac{SNR}{\gamma}=C\cdot\frac{E_b}{N_0}}{=} \log_2\left(1 + C\cdot\frac{E_b}{N_0}\right)$$

Moving forward, all the analysis will be made for constant SNR rather than constant $\frac{E_b}{N_0}$. On Figure 6 you can see similar results for constant SNR rather than $\frac{E_b}{N_0}$. A sparse matrix is still superior to a random one but they both achieve lower capacity than a UNTF. Pay attention that now the capacity achieved by an UNTF is dependent on $\gamma$.

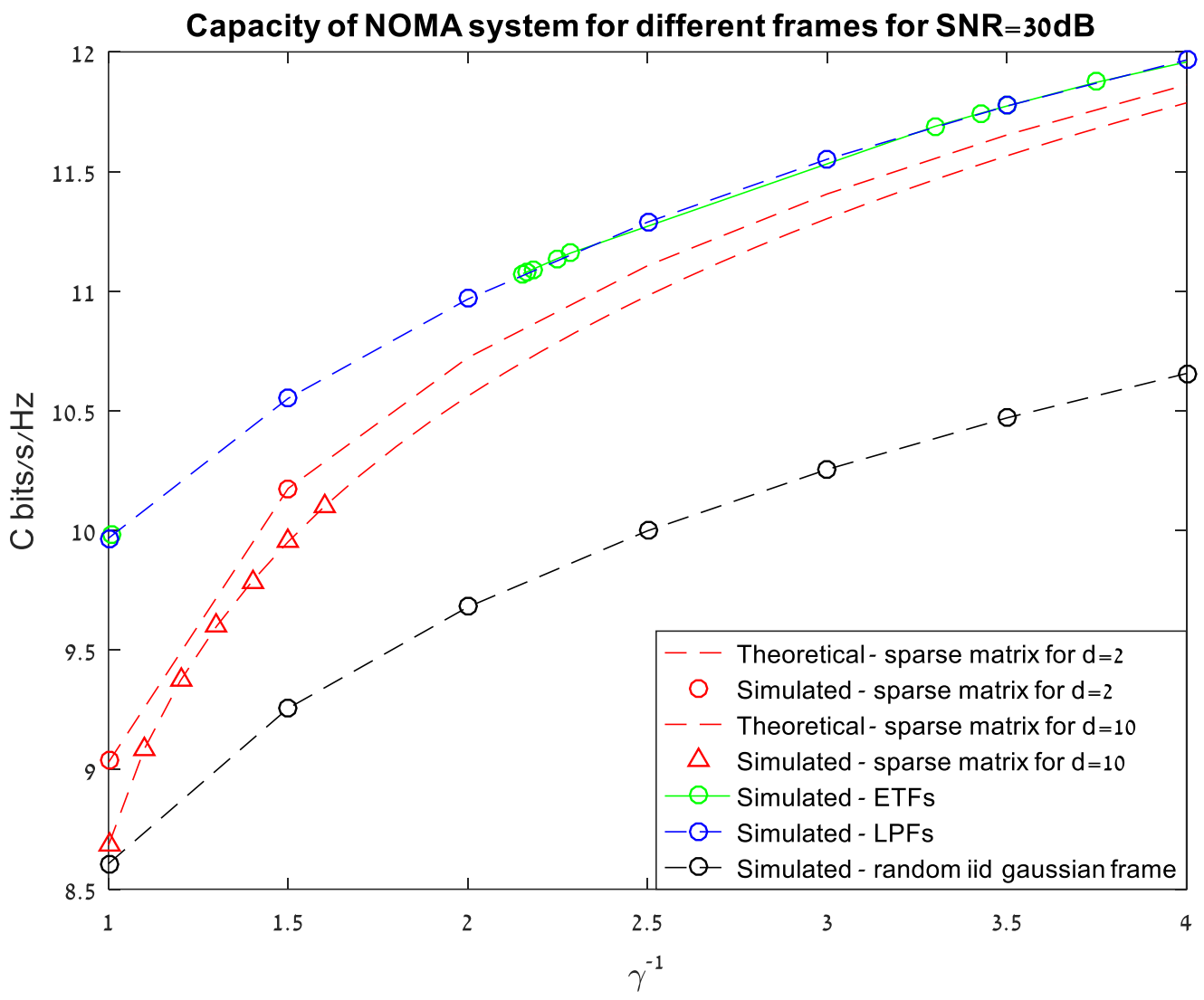

*Figure 6 - Capacity of fully active NOMA system for diferent frames for SNR=30dB*

## 5.2. Partially active NOMA system

We now examine a more realistic setting where only $K$ of all users are active at the same time.

Remember, we defined $\beta \triangleq \frac{K}{M} = \frac{pN}{M}$ as the system active load (number of active users per resource).

### 5.2.1. Results reproduction

We start with reproducing the results given in [1]. The paper simulated the capacity per user for an ETF frame and random i.i.d gaussian frame for different system loads ( $\beta$ ). It didn't specify the SNR for which the results where obtain but from comparing to our results (Figure 7) we can infer they were obtained for $SNR = 0\,dB$ .

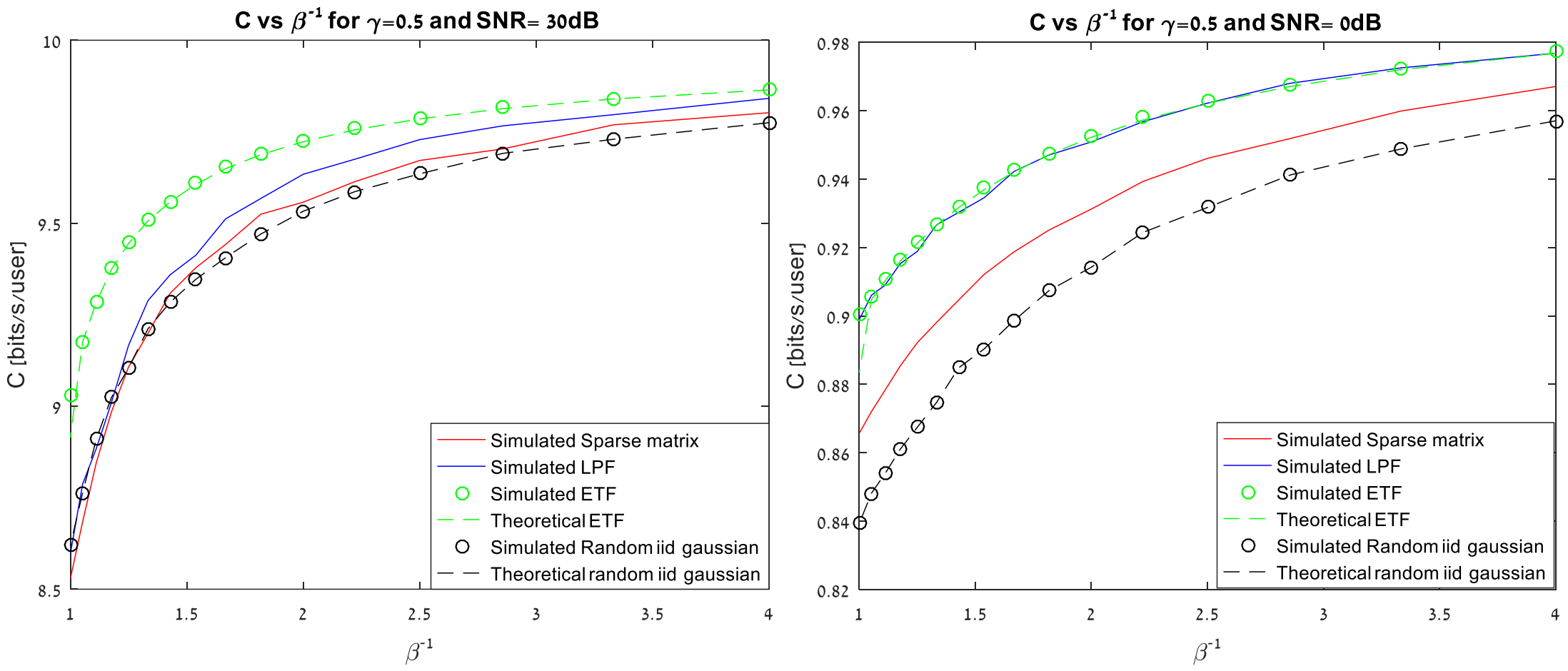

*Figure 7 – capacity comparison for different frames vs the system load (β) for SNR= 0dB and 30dB*

Note we started analyzing the achieved capacity per user first rather than the capacity per resource. We claim there is no point in overloading the system to achieve better spectral efficiency at the expense of the user quality of service. We first must make sure our capacity per user is within reason and then try to make the most of our resources.

As you can see, in a partially active setting there is a difference between using ETFs or UTFs. As proven in [4] the performance of ETFs is superior to all other frames.

### 5.2.2. setup

After much thought we came to the conclusion that this kind of analysis doesn't represent the system well. In paper [1], they chose to keep $\gamma$ constant and change $p$ in order to change $\beta$. However, $p$ is a characteristic of the system and as such should be kept constant within a plot. In order to plot the capacity vs $\beta$, $\gamma$ needs to change to maintain the ratio $p = \beta \cdot \gamma$ constant.

Generating frames with different $\gamma$ ratios is not a problem except for ETFs. As mentioned earlier ETFs exist only for specific M, N pairs i.e. specific $\gamma$ ratios. One of the simplest methods of construction is for families with ration $\gamma = 0.5$. As a result, most papers concentrate their analysis on $\gamma = 0.5$ ETFs exactly like [1]. However, using the Steiner method we were able to construct ETFs with various $\gamma$ ratios and in doing so improve the analysis by keeping the parameter $p$ constant.

While constructing the ETFs we encountered two questions:

1. The Steiner logarithm is quite complex and takes a long time to finish. The larger the frame the longer the running time will be. We wanted to know what is the minimal frame size that will result in enough statistics.
2. ETFs with different $\gamma$ have also different sizes. Meaning M is not constant but rather changes from frame to frame. The question is whether the number of resources (M) effects the capacity achieved or does only the ratio $\gamma$ effect it?

In order to answer both questions, we constructed several ETFs with the same ratio $\gamma = 0.5$ and different number of resources M.
First, for some $p$, we calculated the variance of the achieved capacity after 50 iterations. As you can see from Figure 8 for $M > 50$ we get sufficiently deterministic results.

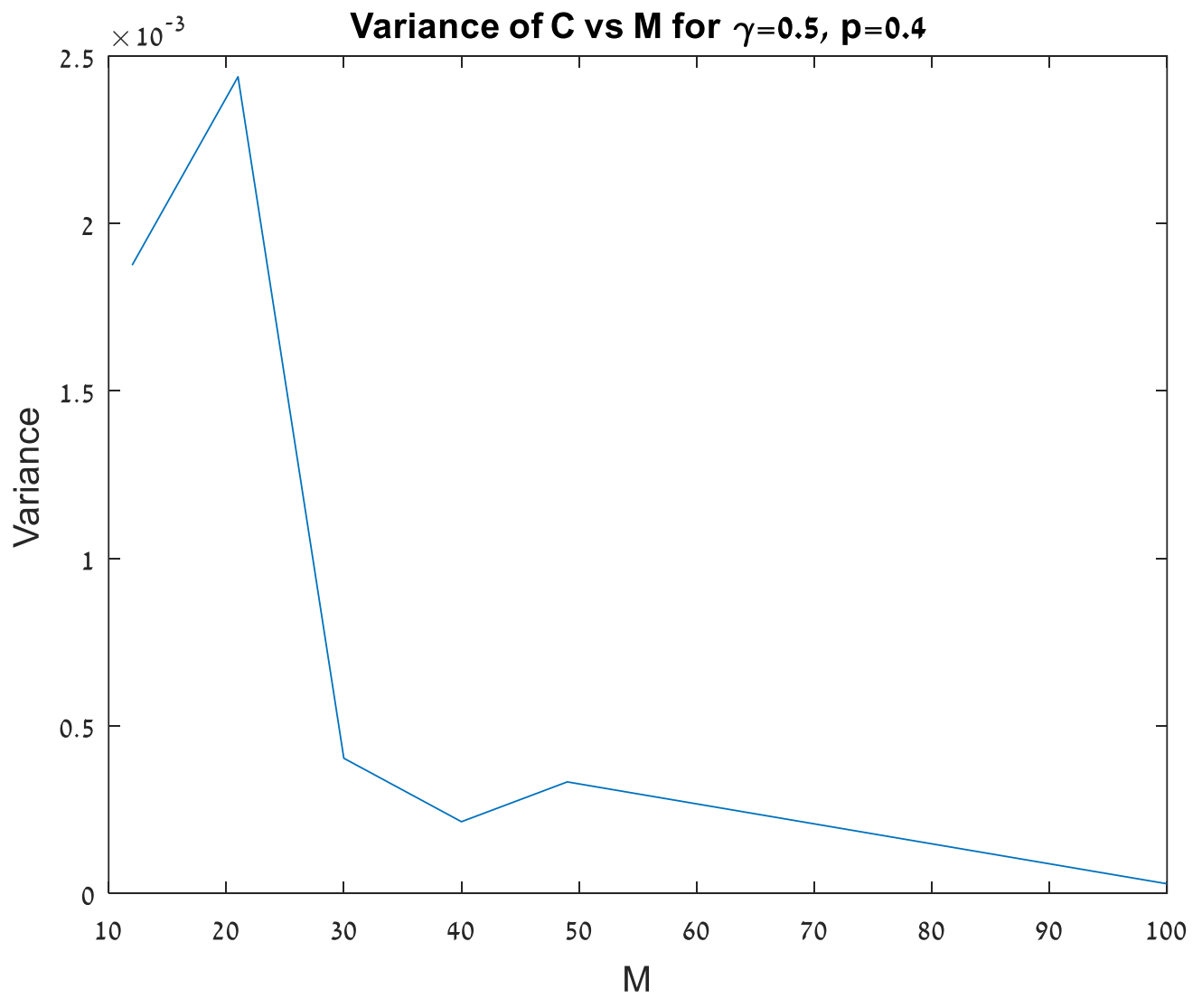

*Figure 8 - Variance of achieved capacity for ETFs with different M values*

Second, for several $p$ 's, we calculated the capacity of the system for the different M's values (averaged over 50 iterations). As you can see in Figure 9 for $M > 50$ M has no effect on the capacity. The affect for smaller M is due to lack of determinism.

In conclusion, from both result we have learned to use only frames of size M bigger than 50 to receive valid results.

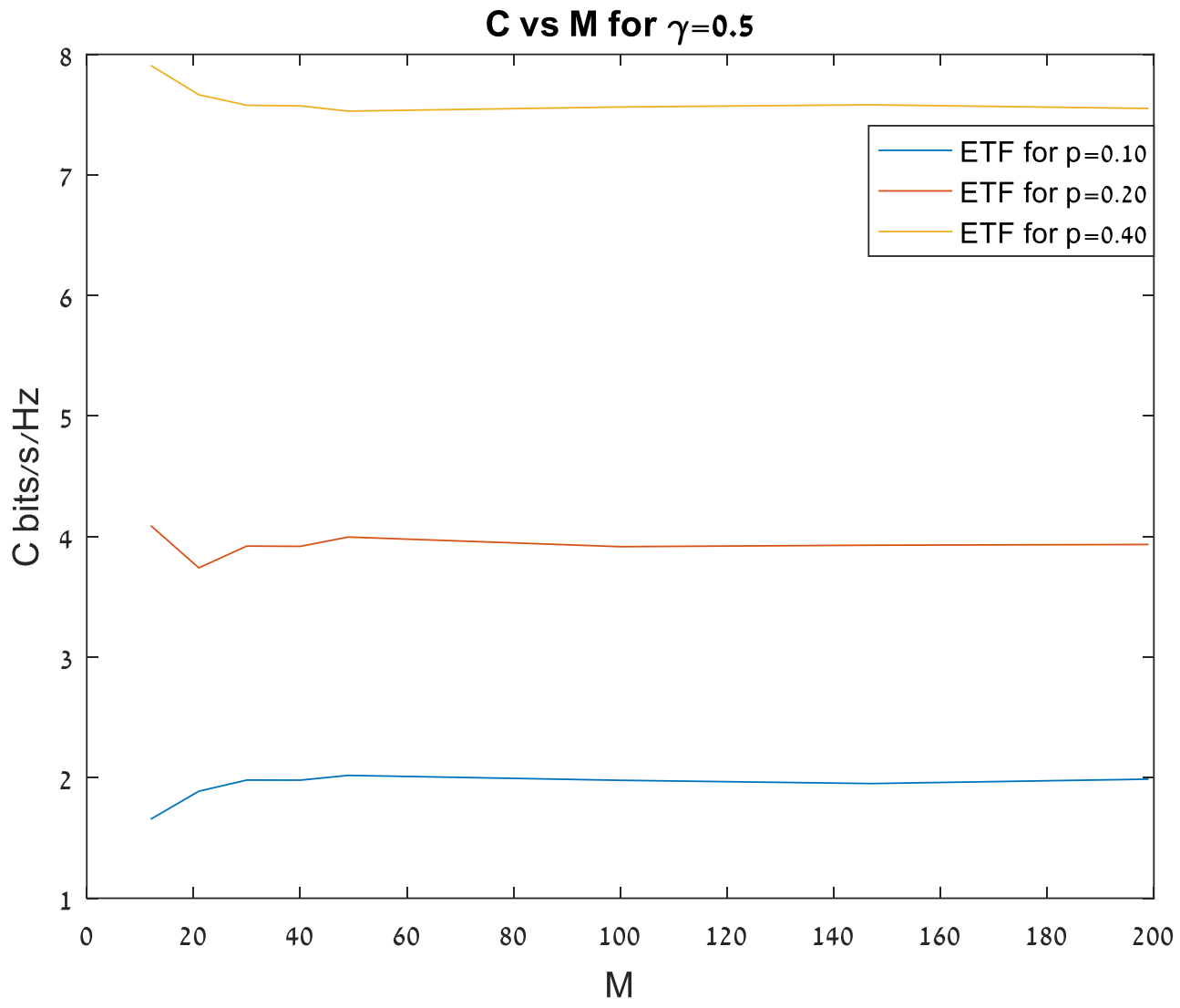

*Figure 9 - capacity for ETFs with differnet M values*

### 5.2.3. Results

Let us review the capacity achieved by the different frames and how $\beta$ affects it. Remember, we keep $p$ constant so $\gamma$ also changes with $\beta$ in order to keep $p = \beta \cdot \gamma$.

In addition, keep in mind that for sparse frames $d\gamma^{-1}$ is limited to natural numbers only, so not all combinations of $\beta, \gamma$ are possible. For example, for $p > 0.66,\ d = 2$ then only one pair of $\beta, \gamma$ is possible:

$$d = 2 \rightarrow \gamma^{-1} = \{1, 1.5, 2...\}$$
$$p > 0.66 \rightarrow \beta = p \cdot \gamma^{-1} = 0.66 \cdot \gamma^{-1} < 1 \rightarrow \gamma = 1 .$$

We will focus on the range $0.5 \le \beta \le 1$.

An interesting observation is that for the same $\beta$, the larger $p$ is the larger the capacity per user (except from random frames which are not affected by $p$). Meaning that for a given final system load (final number of users per resource) we will get larger capacity if the original load was small with high probability of user participance than large load and small probability of participance. The reason for this is that we did the user-resource allocation without knowledge of the final participance status, so the smaller the original load was the smaller the final probability of user to user interference will be.
In addition, its logical that random frames are not affected by the user participance ratio because its random, so for every subframe we choose, the statistical characteristics remain the same.

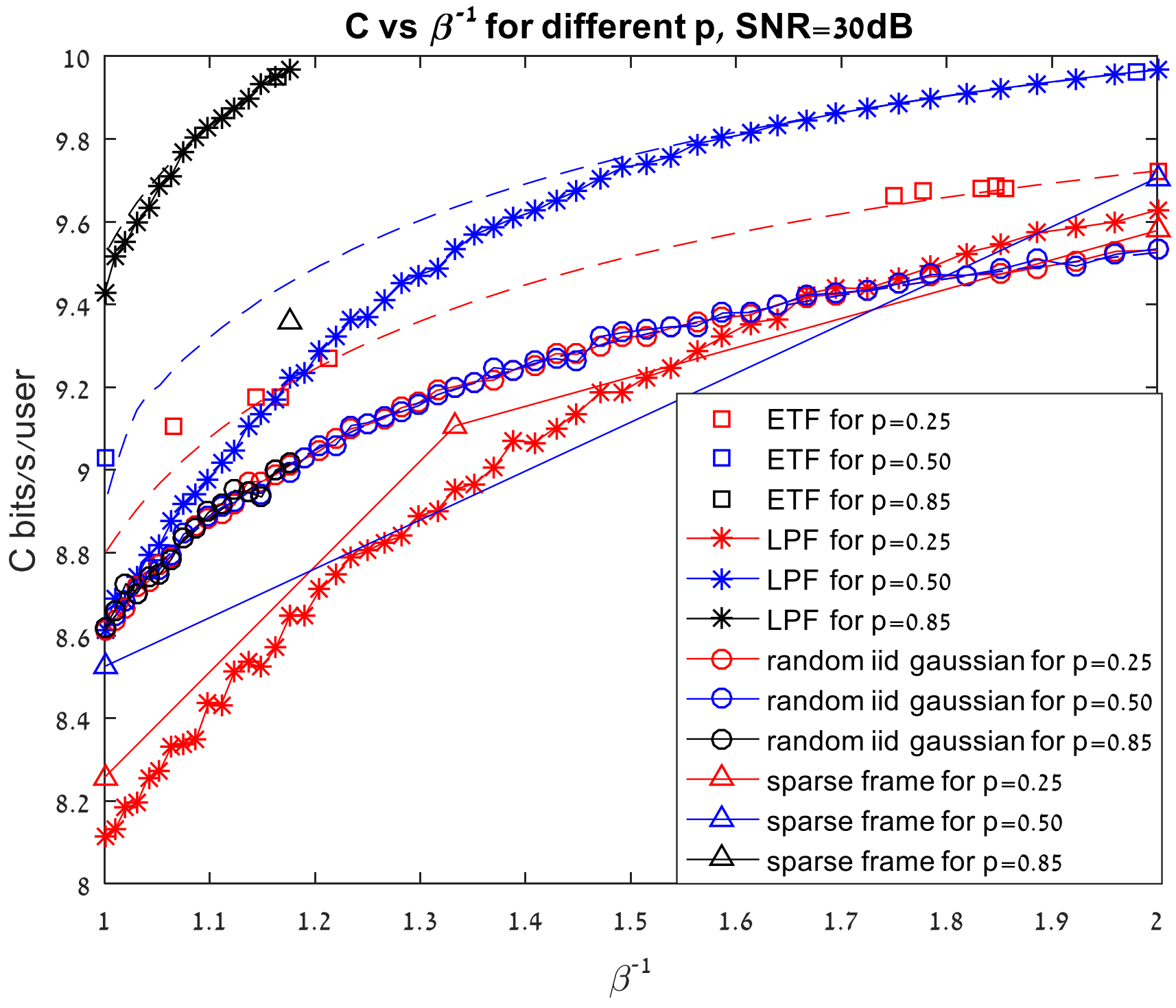

*Figure 10 - Capcity per user vs β for different frame types and p. SNR = 30dB. The sparse frame was constructed with d=2. γ=p/β.*

However, as said before, the user participance ratio $p$ is given for each system and we can't control it. So the question is - **given a constant $p$ what are the best $\beta, \gamma$ pair and frame type we can choose to improve our system's performance?**

Regarding the preferable frame type, we must devide our answer in three:

1. For small $p$ ( $p < 0.25$ ) the random frame gives suprisingly good results which are better from all other frames with the exeption of ETF. However, the ETF advantage is almost neglectable (smaller than half a bit) so its probably preferable to work with randome frames to gain system simplisity.
2. For median $p$ ( $0.25 < p < 0.75$ ) the ETF's superiority is more dominant and may be the favorite choise despite the added complexity to the system. It's notable that the LPF converges to the ETF's capacity for small $\beta$ but this won't be a good operating point as will be shown in Figure 11.
3. For large $p$ ( $p > 0.75$ ) there is almost no difference between the performance of an ETF and a LPF as can be seen from Figure 10 and Figure 6 (for $p = 1$). So, the preferred choice would be a LPF in order to reduce system complexity.

From Figure 10 it's clear that the smaller $\beta$ is, the higher the user capacity. This makes sense since lowering our system's final load decreases the probability of user to user interference. However, decreasing $\beta$ lowers our system efficiency (Figure 11). There is no obvious optimal operating point but rather a trade-off between the two.

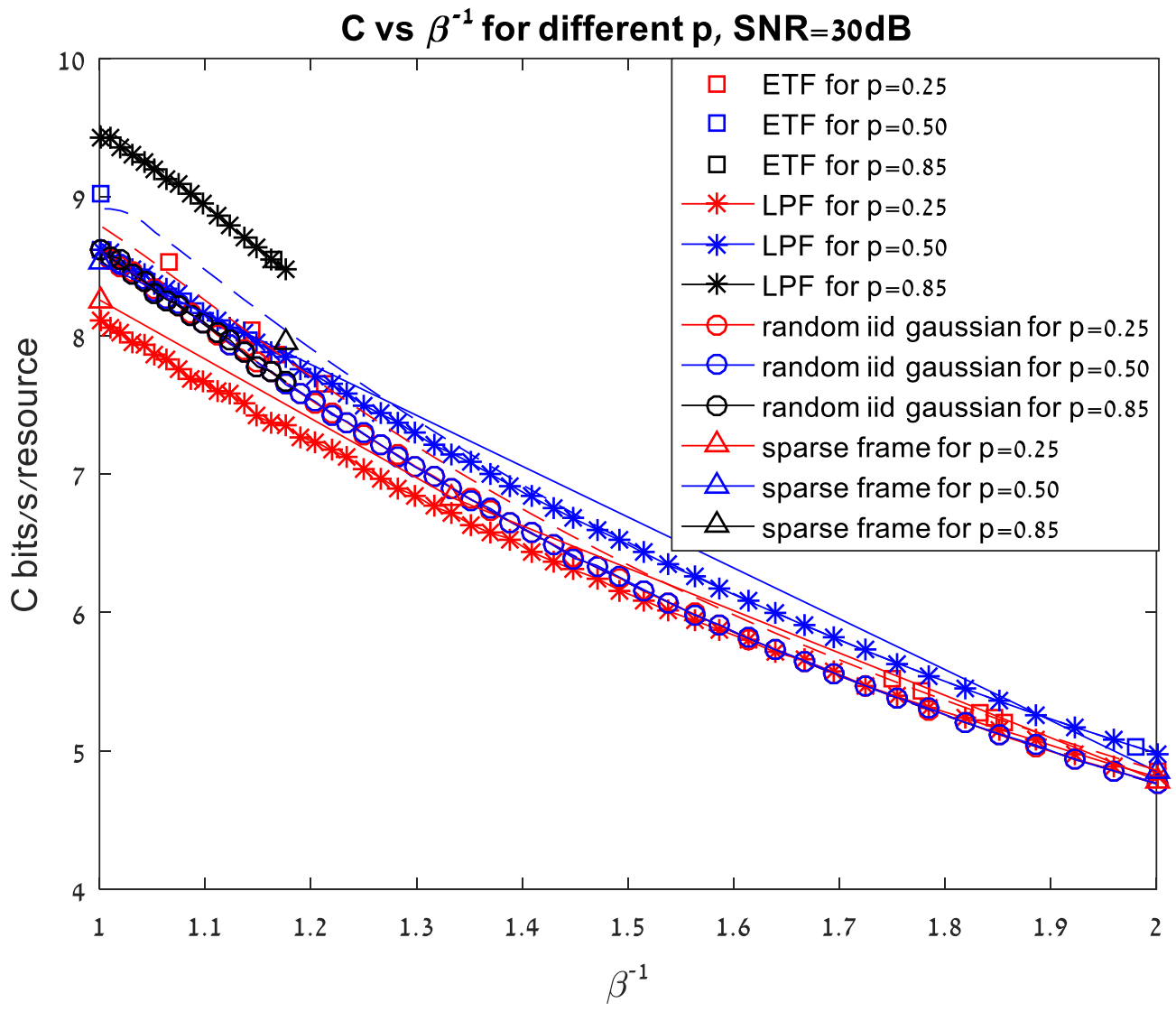

*Figure 11 - capacity per resource vs β for different frame types and p. SNR = 30dB. The sparse frame was constructed with d=2.*

As $\beta \to 1$ the eigenvalues grows smaller achieving zero at $\beta = 1$ , which will cause the capacity per user to decrease significantly (Equation 3). If the slope of the capacity per user vs $\beta^{-1}$ is higher than one (higher than the slope caused by the multiplication with $\beta$ ; Equation 5) , we should have gotten a maximum point at $\beta \neq 1$.

Meaning, a zeroed eigenvalue could have caused a maximum point at $\beta \neq 1$ for the capacity per resource and as a result, an obvious optimal operating point.

Which raises the question – why does the capacity per resource decreases monotonically with $\beta^{-1}$?

The hypothesis was that the identity matrix in the equation was preventing the capacity per user to decrease significantly when $\beta \to 1$. If we omit the identity matrix we get:

$$C\left[bits/s/user\right] = \frac{1}{K}\sum_{i=1}^{K}\log_2\left(SNR\cdot\lambda_i\right) = \log_2\left(SNR\right) + \frac{1}{K}\sum_{i=1}^{K}\log_2\left(\lambda_i\right) =$$

$$\log_2\left(SNR\right) + \underbrace{\log_2\left(\left(\prod_{i=1}^{K}\lambda_i\right)^{1/K}\right)}_{LGM} = \log_2\left(SNR\right) + LGM\left(\beta\right) \xrightarrow[\beta\to 1]{} -\infty$$

$$C\left[bits/s/resource\right] = \beta\cdot C\left[bits/s/user\right] = \beta\cdot\log_2\left(SNR\right) + \beta\cdot LGM\left(\beta\right)$$

LGM, stands for Log Geometric Mean.

Meaning, if we omit I, the capacity per user goes to $-\infty$ when $\beta \to 1$. This will cause the capacity per resource to increase with $\beta^{-1}$ at $\beta \to 1$ and enable us to deduce an optimal operating point.

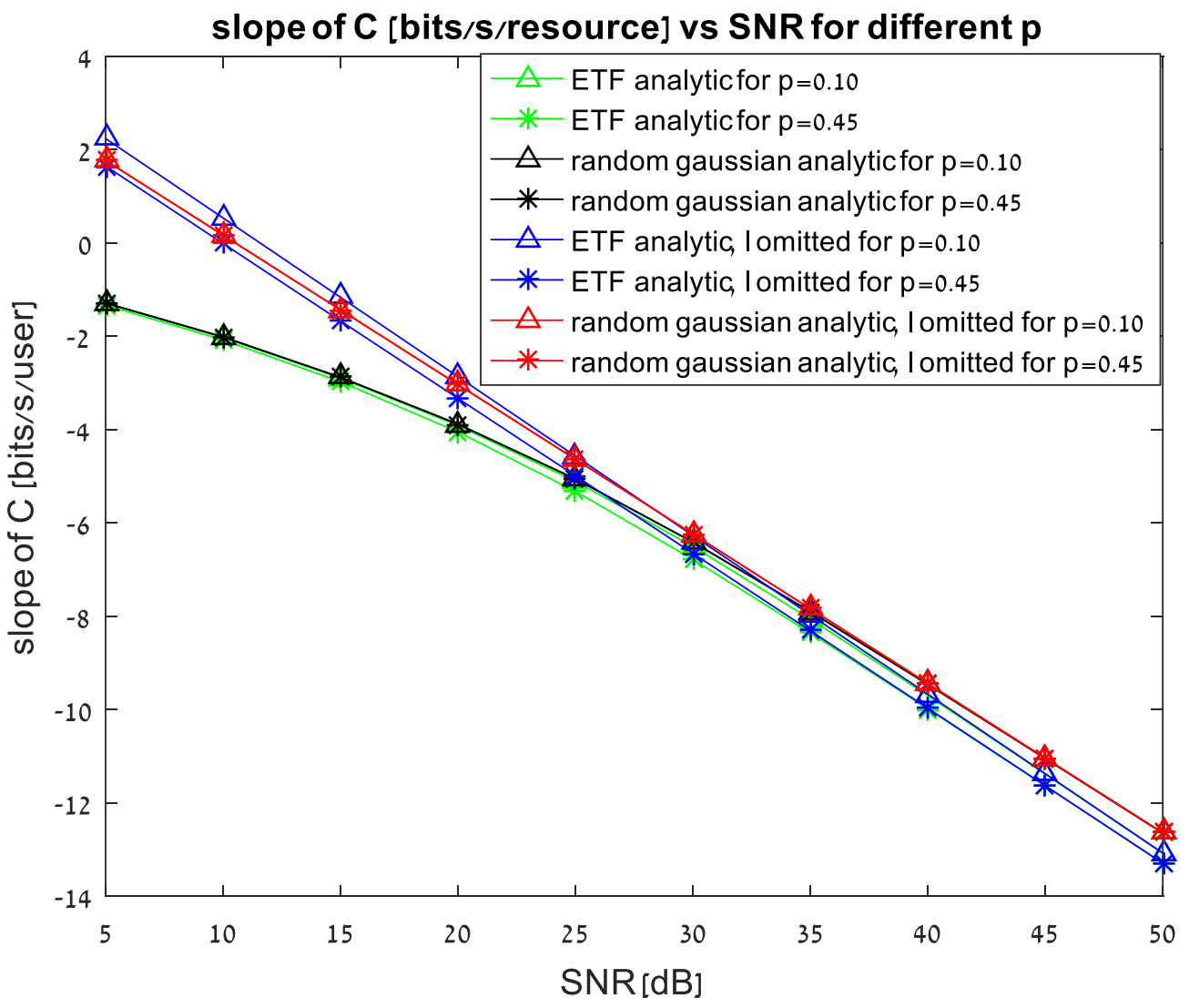

*Figure 12 – slope at β=0.95 of graph, capcity per resource vs β^-1, as a function of SNR for different frame types and p*

From Figure 12 you can see our hypothesis was correct. The capacity per resource will never have a maximum point at $\beta \neq 1$ (the slope is always lower than 1). However, if we neglect the identity matrix, which is in fact examining the practical capacity as defined in Equation 4, we see it is possible but depends on the SNR. As an example, you can view Figure 13.

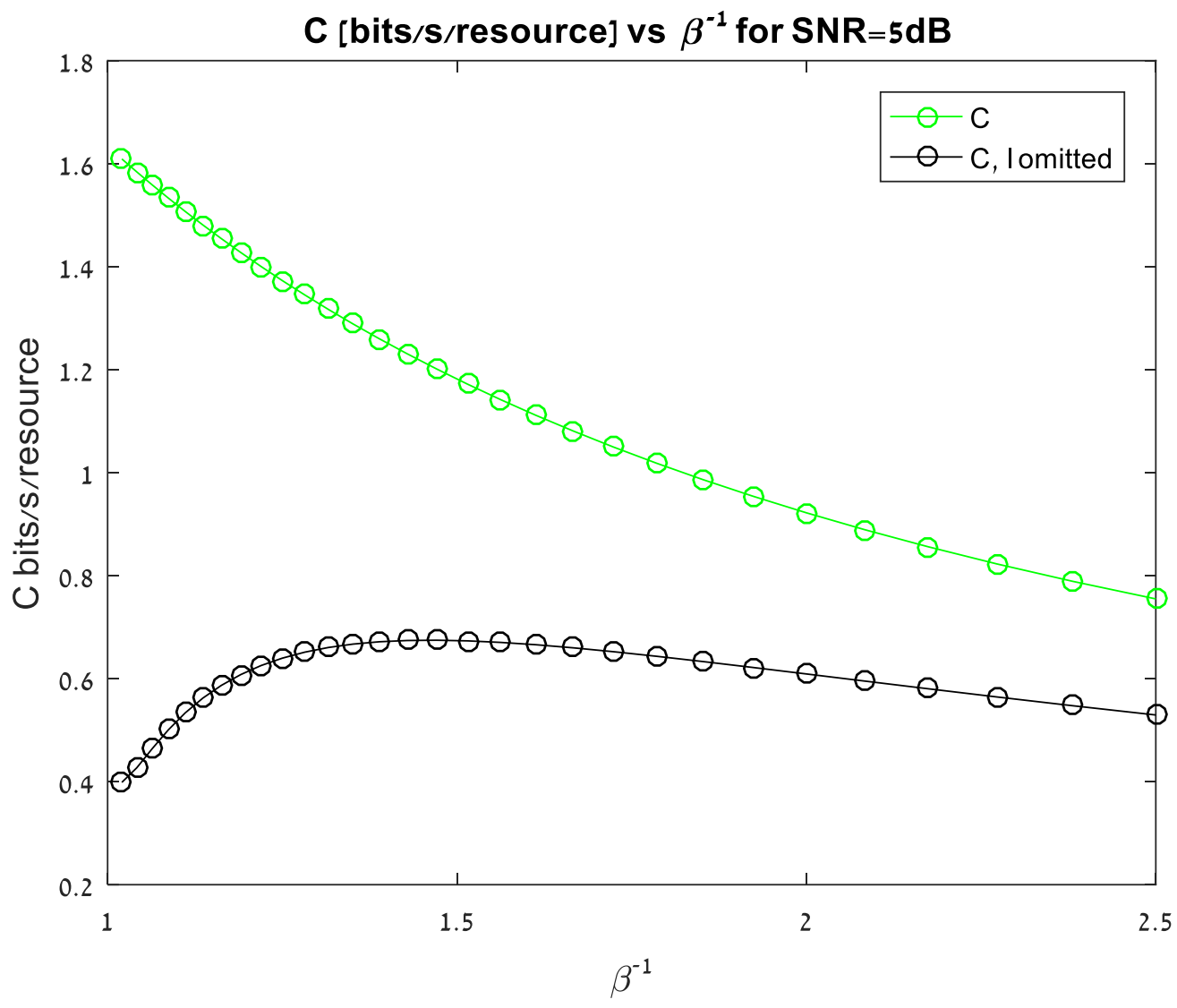

*Figure 13 - example of maximum point at β≠1 when we omit I. SNR = 5dB*

### *Practical capacity*

Let us examine the practical capacity rate performance of the different frames. We will start by looking on the LGM first.

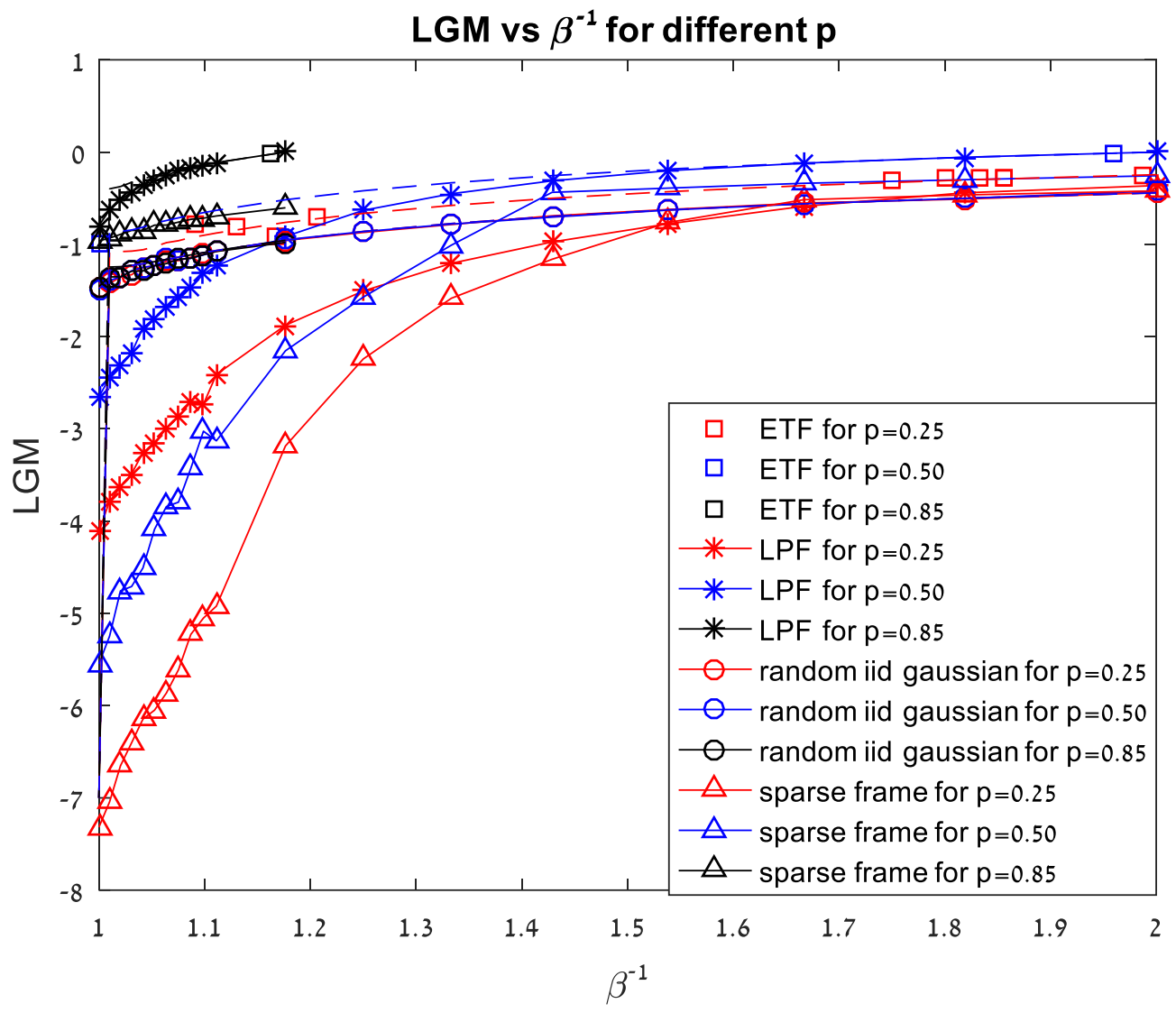

*Figure14 - LGM performance for different frame types and p VS β ^-1*

From Figure14 we can see that the eigenvalues of ETFs and random frames remain relatively steady even for $\beta = 1$. The theoretical calculation, however, equals $-\infty$ at $\beta = 1$ because apparently the possibility of getting a zero eigenvalues exists. Only a slight movement from $\beta = 1$ ensures we get eigenvalues larger than zero.

The case for LPF and sparse frames is totally different. It's clear that the eigenvalues of the frames grow smaller rapidly and are significantly affected by the size of $\beta$. This implies the practical capacity's maximum point for these two types of frames will be achieved at smaller $\beta$ than for ETF and random frames.

Note that Figure14 can also represent the practical capacity per user at SNR=1dB as the SNR only adds a constant to all graphs. For example, you can look at Figure 15 and see its identical to Figure14 with an added constant.

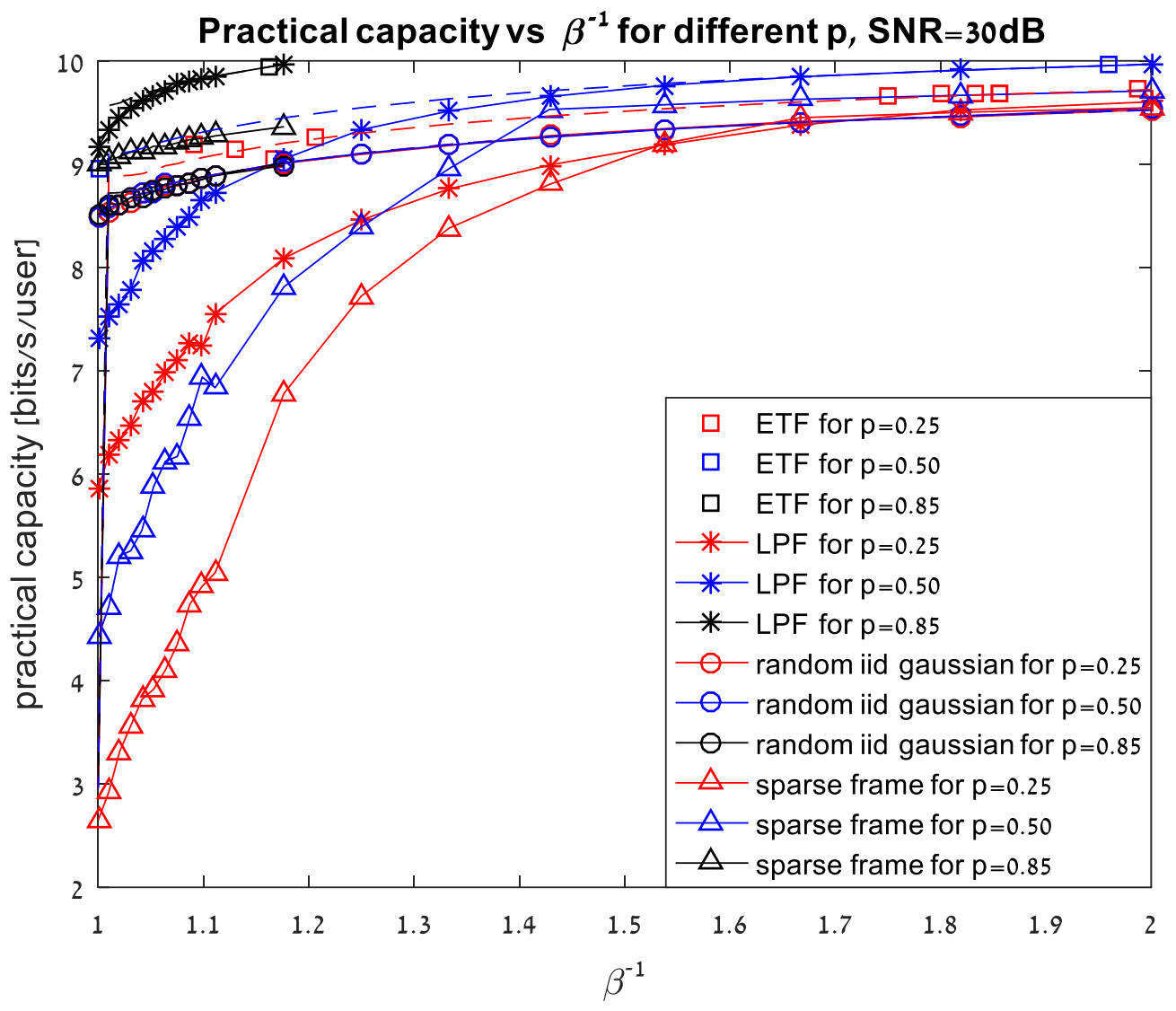

*Figure 15 – practical capacity rate per user for differnet frame type VS β^-1 , for SNR=30dB*

Now, examining the practical capacity per resource in Figure16 , we can see it coincides with our previous conclusions. The optimal point for ETFs and random frames is indeed achieved at much higher $\beta$ than LPFs and sparse frames.

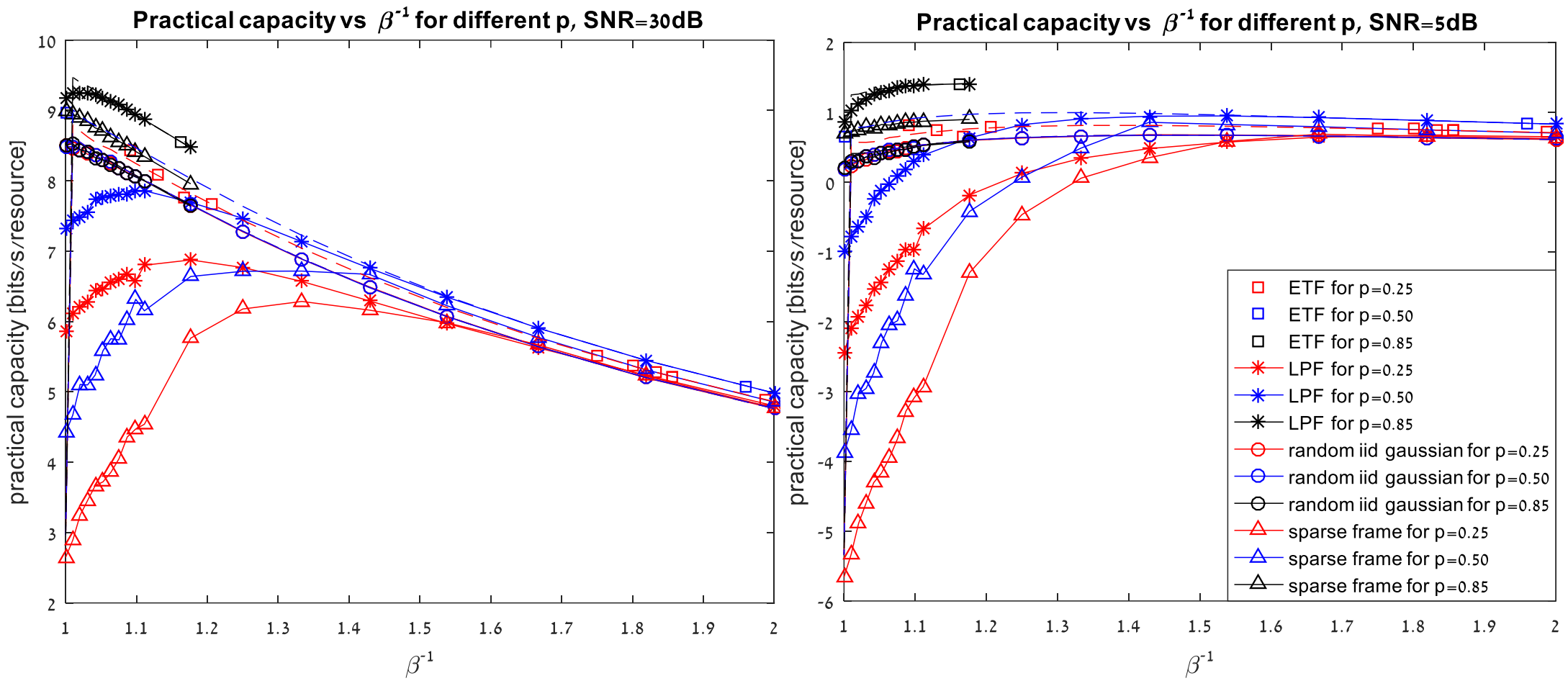

*Figure16 – practical capacity per resource for different frame types VS β^-1, for different SNRs*

## *SNR effect*

Another factor to take into consideration is the SNR. Larger SNRs drive the optimal point closer to 1 for all frames. For ETFs and random frames the drift towards 1 is very quick, and even when the optimum is far from 1, the capacity difference is neglectable. Meaning the optimal point operating point will always be a nudge from 1. For the other frames the drift is slower, and the optimal operating point will be for some $\beta < 1$ as can be seen in Figure 17.

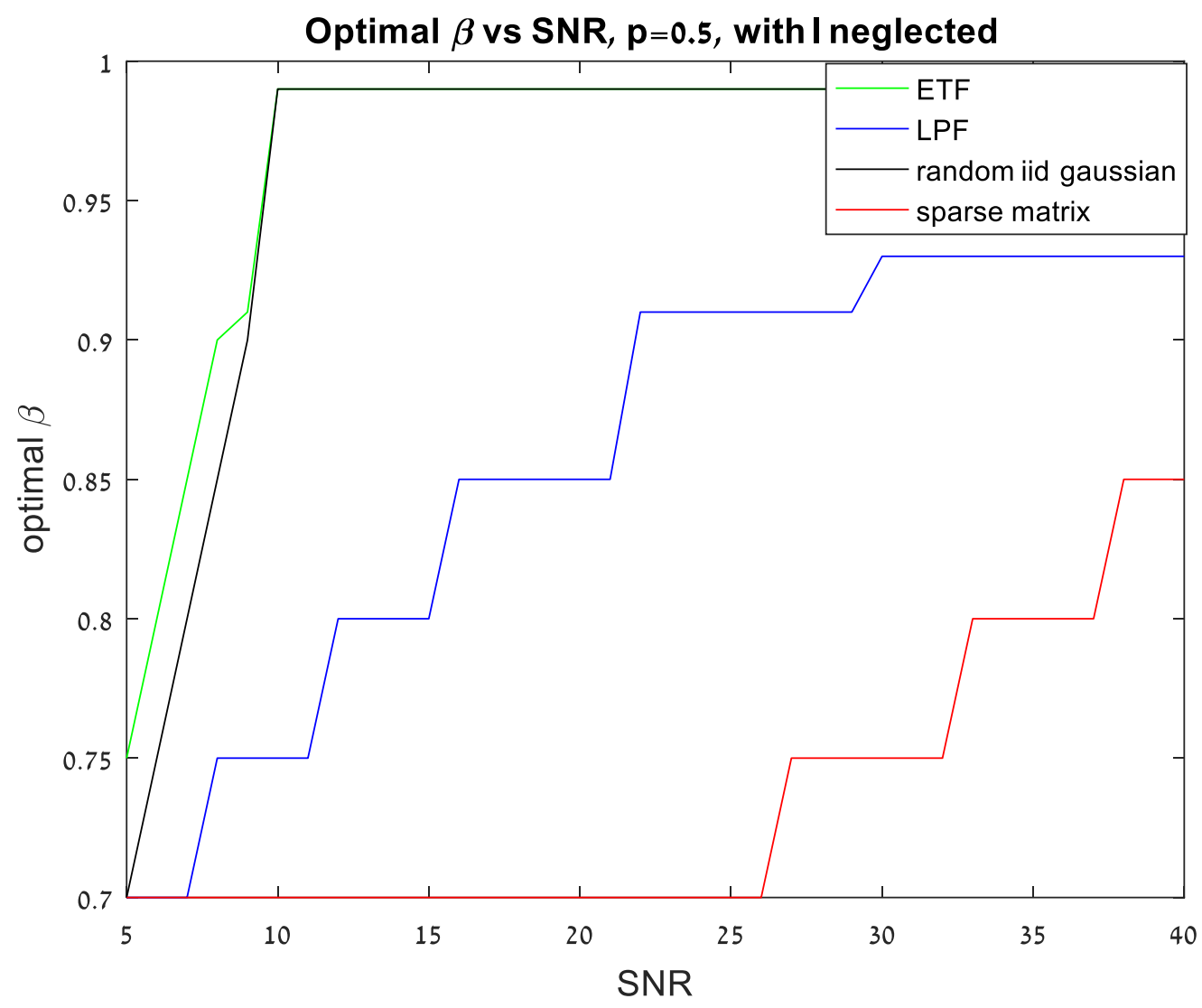

*Figure 17 –practical capacity rate's optimal β drift toward 1 with SNR for different frame types*

Looking at the optimal $\beta$ for capacity per resource, Figure 18 , we can see the effect of the identity matrix which canceled the fading towards 1 causing the optimal $\beta$ to be always a nudge from 1 for all frame types and all SNR. Meaning the SNR has no effect on the capacity per resource's optimal operating point.

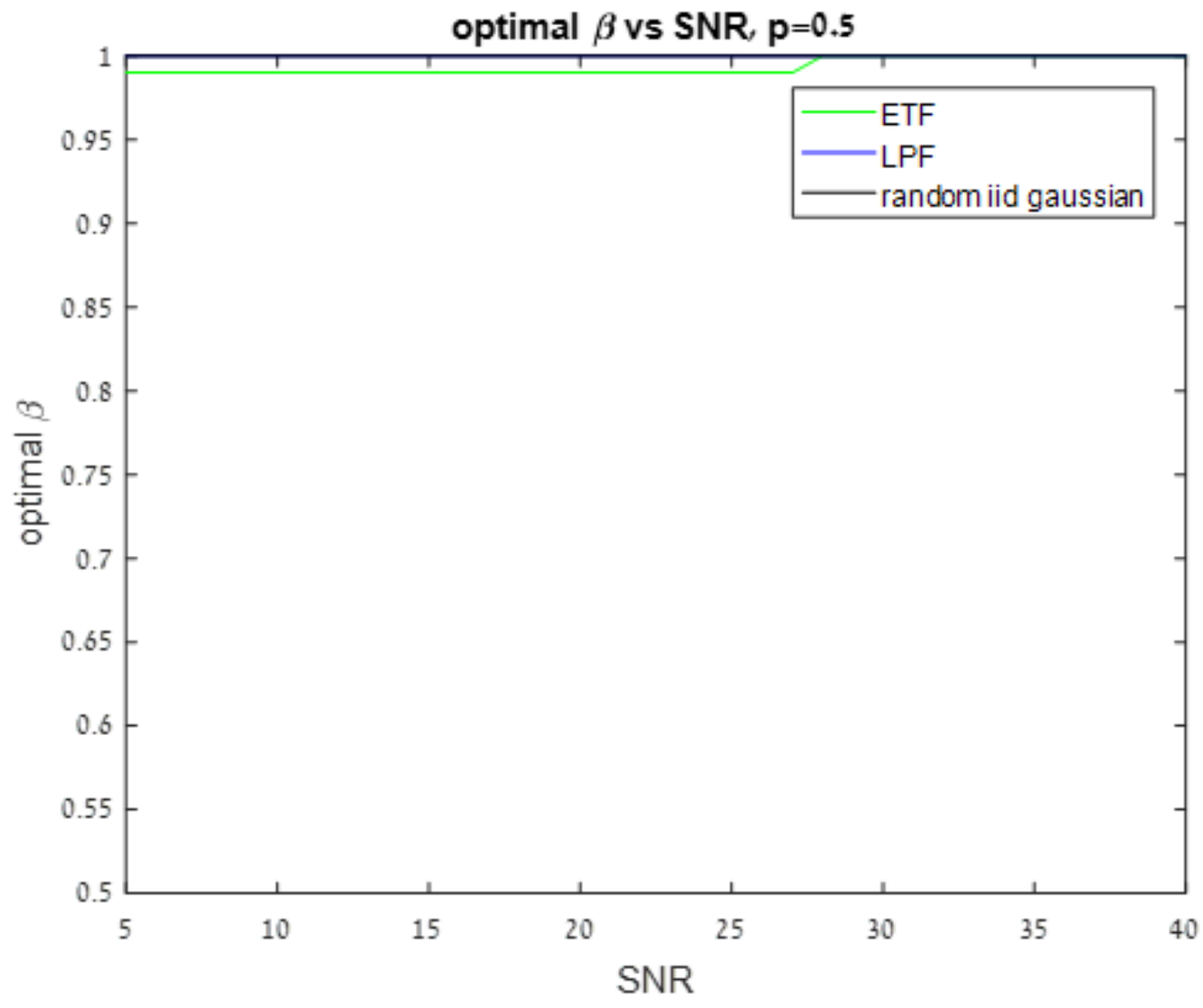

*Figure 18 - capacity's optimal β drift toward 1 with SNR for different frame types*

# 6. Discussion and Future Work

We started this paper by briefly examining the special case of fully active NOMA system ( $p=1$). We saw the superiority of sparse frames over random ones but that the largest capacity was achieved by unit norm tight frames of any kind.

We proceeded to the general case of partially active system were only part of the users are active simultaneously. We asked ourselves, given a constant $p$ ,what are the best $\beta,\gamma$ pair and frame type we can choose to improve our system's performance?

The performance analysis was divided into two cases – (1) Capacity analysis and (2) practical capacity analysis, a realist maximal rate achieved by simple coding schemes. The latter helped understand the influence of the eigenvalues and the SNR on the system.

For the capacity analysis we saw there is a constant tradeoff between system efficiency and user capacity when choosing the best $\beta,\gamma$ pair. Since the user capacity reaches a plateau and can't be improved significantly, the best choice for $\beta$ would be the one maximizing the spectral efficiency which is a nudge from 1 or even 1 for large SNRs.
Regarding the best frame type, the answer depends on $p$ :

1. For small $p$ ( $p<0.25$) random frames give good results and even the ETF advantage is neglectable. So it all depends on generation convenience.
2. For medium $p$ ($0.25<p<0.75$) the ETF's superiority is dominant and is the obvious choise.
3. For large $p$ ( $p>0.75$) , including the special case of fully active, there is almost no difference between the performance of an ETF and a LPF . So, the preferred choice would be a LPF in order to reduce system complexity.

Analizing the practical capacicty helped us visulaize the affect of the eigenvalues and the SNR . We saw, for all frame types, that the user capacity goes to $-\infty$ at $\beta=1$ due to the probability of a zeroed eigenvalue. However, while the eigenvalues of ETFs and random frames recuperate rapidly, the eigenvalues of LPFs and sparse frames remain small (and so does the user capacity) for a large range of $\beta$ s.
This is the reason why the spectral efficiency's optimal $\beta$ is:

1. a nudge from 1 for ETFs and random frames.
2. Depends on the SNR for LPFs and sparse frames, grows closer to 1 with the increase of SNR.

when choosing the preferred frame type the choice is between ETF and random frame for two reasons:

1. They achieve larger practical capacity than LPF and sparse frame.
2. their operating point is not dependent on the SNR.

ETF always achieves the best results, but its advantage diminishes with the decrease of $p$ .

In conclusion setting the system so that $\beta$ is a nudge from 1 is the optimal choice. If sufficiently complex coding schemes can be employed in order to reach capacity, random frames, ETFs or LPFs can be used for resource-user mapping, depending on the user participance ratio. Further work can be dedicated to finding a better UNTF that will reduce the advantage of ETF at medium size $p$. In addition, more types of difference sets can be explored to broaden the range of ETF's dimensions.

# 8. Appendixes

## 8.1. Matlab code – sparse matrix generation

```
function [F] = create_zaidel_matrix(M,betta,d)
tries = 5;
N = round(betta*M); %num of users
F = zeros(M,N);

ms = repmat((1:M),1,round(betta*d));
ns = repmat((1:N),1,d);

fail=1;
for i = 1:N*d
    done_before = 1;
    rand_angle = 2*pi*rand;
    rand_number = cos(rand_angle)+1i*sin(rand_angle);
    for t = 1:tries
        m_idx = randi(length(ms));
        n_idx = randi(length(ns));

        m = ms(m_idx);
        n = ns(n_idx);

        if F(m,n)==0
            fail=0;
            break;
        end
    end
    if fail == 1
        F = NaN;
        break
    end
    F(m,n) = rand_number;
    ms(m_idx) = [];
    ns(n_idx) = [];
end
end
```

## 8.2. Matlab code – LPF generation

```
function [F] = generate_LPF(M,N)

F = dftmtx(N);
F = F(1:M,:);
h = F(:,1);
F = sqrt(1/abs(h'*h))*F;

check_utf = abs(F'*F);
if all(all(F*F' - N/M*eye(M)<0.0001)) && all(diag(check_utf)-
1<0.0001)
    fprintf('Matrix F is UTF\n');
end
```

## 8.3. Python code - Steiner ETF construction

```
import numpy as np
import itertools
```

```
from concurrent import futures

def main():
    M = 70
    beta = 3.3
    N = beta * M
    k = 3
    v = 21
    r = 10
    b = M

    elements = list(itertools.combinations(range(v),k))
    A = np.zeros((int(b), v))
    elements = random.sample(elements, len(elements))
    answer = set_block(A.copy(), 0, elements, v-1,r)
    print(answer)

    current_col = 0
    for col in range(v):
        rows = answer[col]
        if rows == ['OK']:
            continue
        if col == 0:
            block_start = 0
        else:
            block_start = np.argwhere(A[:, col - 1] == 1)[-1][0] + 1
        for i in range(len(rows)):
            row = rows[i]
            for j in row:
                A[block_start + i][j] = 1
    print(A)
    np.savetxt('matrix.csv', A, delimiter=',')




def set_block(A,col,elements,max_col,r):

    combin = [x for x in elements if x[0] == col]

    if not combin:
        if col == max_col:
            return [['OK']]
        elif r - np.sum(A[:, col]) == 0:
            block_answer = set_block(A.copy(), col + 1,
elements.copy(), max_col, r)
            return [['OK']]+block_answer

    [elements.remove(x) for x in combin]

    for i in range(len(combin)):
        if (col == 2) and (i <= 278):
            continue
        if col<=3:
            print([col,i])
        temp_A = A.copy()
        temp_elements = elements.copy()
        answer = set_row(combin.copy(), row=i, deep=0, max_deep=r-
np.sum(A[:, col])-1)
        if answer:
```

```
            if col == max_col:
                return answer
            else:
                if col == 0:
                    block_start = 0
                else:
                    block_start = np.argwhere(temp_A[:, col - 1] ==
1)[-1][0]+1
                for a in range(len(answer)):
                    element = answer[a]
                    two_comb = list(itertools.combinations(element,
2))

                    for j in element:
                        temp_A[block_start+a][j] = 1
                    for j in range(len(temp_elements) - 1, -1,
                                   -1):  # every combination of two
elements must appear in only on row
                        for c in two_comb:
                            # print([elements[j],c])
                            if all(val in temp_elements[j] for val in
c):
                                temp_elements.pop(j)
                                break
                block_answer = set_block(temp_A.copy(),col+1,
temp_elements.copy(), max_col,r)
                if block_answer:
                    return [answer] + block_answer

    return []



def set_row(A,row,deep,max_deep):
    element = A[row]
    two_comb = list(itertools.combinations(element,2))

    for i in range(len(A)-1, -1, -1):         # every combination of
two elements must appear in only on row
        for c in two_comb:
            if all(val in A[i] for val in c):
                A.pop(i)
                break
    if deep == max_deep:
        return [element]
    elif A:                        # if A is not empty
        for i in range(len(A)):
            response = set_row(A.copy(),i,deep+1,max_deep)
            if response:
                return [element]+response
    return []

if __name__ == '__main__':
    main()
```

## 8.4. Python code - Steiner ETF construction using parallel computing

```
import numpy as np
import itertools
from concurrent import futures

def main():
    M = 70
    beta = 3.3
    N = beta * M
    k = 3
    v = 21
    r = 10
    b = M

    elements = list(itertools.combinations(range(v),k))
    A = np.zeros((int(b), v))
    elements = random.sample(elements, len(elements))
    answer = set_block(A.copy(), 0, elements, v-1,r)
    print(answer)

    current_col = 0
    for col in range(v):
        rows = answer[col]
        if rows == ['OK']:
            continue
        if col == 0:
            block_start = 0
        else:
            block_start = np.argwhere(A[:, col - 1] == 1)[-1][0] + 1
        for i in range(len(rows)):
            row = rows[i]
            for j in row:
                A[block_start + i][j] = 1
    print(A)
    np.savetxt('matrix.csv', A, delimiter=',')

def set_block(A,col,elements,max_col,r):

    combin = [(x,y,z) for (x,y,z) in elements if x == col]

    if not combin:
        if col == max_col:
            return [['OK']]
        elif r - np.sum(A[:, col]) == 0:
            block_answer = set_block(A.copy(), col + 1,
elements.copy(), max_col, r)
            return [['OK']]+block_answer

    [elements.remove(x) for x in combin]
    if col == 4:
        with futures.ProcessPoolExecutor() as pool:
            for response in list(pool.map(content,
range(len(combin)), itertools.repeat(col),
itertools.repeat(A),itertools.repeat(elements),itertools.repeat(combi
n),itertools.repeat(r),itertools.repeat(max_col))):
                if response:
                    return response
    else:
        for i in range(len(combin)):
            response = content(i, col, A,elements,combin,r,max_col)
            if response:
```

```
                return response

    return []

def content(i,col,A,elements,combin,r,max_col):
    if col <= 4:
        print([col, i])

    temp_A = A.copy()
    temp_elements = elements.copy()
    answer = set_row(combin.copy(), row=i, deep=0, max_deep=r -
np.sum(A[:, col]) - 1)
    if answer:

        if col == max_col:
            return answer
        else:
            if col == 0:
                block_start = 0
            else:
                block_start = np.argwhere(temp_A[:, col - 1] == 1)[-
1][0] + 1
            for a in range(len(answer)):
                element = answer[a]
                two_comb = list(itertools.combinations(element, 2))

                for j in element:
                    temp_A[block_start + a][j] = 1
                for j in range(len(temp_elements) - 1, -1,
                               -1):  # every combination of two
elements must appear in only on row
                    for c in two_comb:
                        # print([elements[j],c])
                        if all(val in temp_elements[j] for val in c):
                            temp_elements.pop(j)
                            break
            block_answer = set_block(temp_A.copy(), col + 1,
temp_elements.copy(), max_col, r)
            if block_answer:
                return [answer] + block_answer
    return []

def set_row(A,row,deep,max_deep):
    element = A[row]
    two_comb = list(itertools.combinations(element,2))

    for i in range(len(A)-1, -1, -1):         # every combination of
two elements must appear in only on row
        for c in two_comb:
            if all(val in A[i] for val in c):
                A.pop(i)
                break
    if deep == max_deep:
        return [element]
    elif A:                        # if A is not empty
        for i in range(len(A)):
            response = set_row(A.copy(),i,deep+1,max_deep)
            if response:
                return [element]+response
    return []
```

```
if __name__ == '__main__':
    main()
```

### 8.5. Matlab code - Gaussian i.i.d. Frame construction

```
function [F] = generate_Gau_iid(M,N)

F = normrnd(0,1/M,M,N);

norm = 1./sqrt(diag(abs(F'*F)));
norm = repmat(norm.',M,1);
F = norm.*F;
check_utf = abs(F'*F);
if  all(diag(check_utf)-1<0.0001)
    fprintf('Matrix F is normalized\n');
end
```